# Title: Images of a first order spin-reorientation phase transition in a metallic kagome ferromagnet

**Authors**

Kevin Heritage[1], Ben Bryant[2]§, Laura A. Fenner[3], Andrew S. Wills[3], Gabriel Aeppli[4,5,6], and Yeong-Ah Soh[1]*

**Affiliations**

[1]Department of Materials, Imperial College London, SW7 2AZ, UK.

[2]London Centre for Nanotechnology, University College London, WC1H 0AH, UK.

[3]Department of Chemistry, University College London, London, WC1H 0AJ, UK.

[4]Laboratory for Solid State Physics, ETH Zurich, Zurich, CH-8093, Switzerland.

[5]Paul Scherrer Institut, Forschungsstrasse 111, 5232 Villigen PSI, Switzerland.

[6]Institut de Physique, Ecole Polytechnique Fédérale de Lausanne (EPFL), CH-1015 Lausanne, Switzerland.

§Current address: High Field Magnet Laboratory (HFML-EMFL), Radboud University, 6525 ED Nijmegen, Netherlands.

*Correspondence to: yeongahsoh@gmail.com

**Abstract:** First order phase transitions, where one phase replaces another by virtue of a simple crossing of free energies, are best known between solids, liquids and vapours, but they also occur in a wide range of other contexts, including even elemental magnets. The key challenges are to establish whether a phase transition is indeed first order, and then to determine how the new phase emerges because this will determine thermodynamic and electronic properties. Here we meet both challenges for the spin reorientation transition in the topological metallic ferromagnet $Fe_3Sn_2$. Our magnetometry and variable temperature magnetic force microscopy experiments reveal that, analogous to the liquid-gas transition in the temperature-pressure plane, this transition is centred on a first order line terminating in a critical end point in the field-temperature plane. We directly image the nucleation and growth associated with the transition and show that the new phase emerges at the most convoluted magnetic domain walls for the high temperature phase and then moves to self organize at the domain centres of the high temperature phase. The dense domain patterns and phase coexistence imply a complex inhomogeneous electronic structure, which will yield anomalous contributions to the magnetic field- and temperature-dependent electrical conductivity.

**One Sentence Summary:** A magnet imaged in exquisite detail shows how the material transforms from one magnetic state to another upon cooling when the transformation is discontinuous, and in particular how the new state emerges and expands.

## Introduction

Ferromagnetic and antiferromagnetic materials often display a preferential orientation of the magnetic moments along magnetic "easy" axes, which in single crystals coincide with high symmetry crystallographic directions. This magneto-crystalline anisotropy arises from spin-orbit coupling, which is a microscopic, electronic property, as well as from long-range dipolar interactions. While in most magnets the easy axes do not change within the magnetically ordered state, some systems[1,2,3], including even elemental chromium[4], undergo spin-reorientation, where the preferred axes change as the temperature is varied since the magnetic anisotropy is temperature-dependent. Such spin-reorientation transitions can be classified as first or second-order depending on whether the magnetic easy axis changes abruptly during the spin reorientation, with domains of the new phase nucleating in the presence of the initial phase, or continuously with the whole system undergoing a homogeneous transition. Understanding such transitions is important not only from the viewpoint of fundamental statistical physics, but also because the static domain patterns characteristic of first order transitions can have very substantial effects on electrical properties which one might wish to exploit for spintronics.

While spin reorientation transitions have been studied extensively[5], their order is open to debate, as demonstrated for non-stoichiometric $DyFe_{11}Ti$ bulk crystals[6] where the sensitivity of the spin reorientation temperature to sample composition leads to an averaging of the transition temperature in bulk crystals that make the transition observed in magnetometry appear to be continuous and second-order. Images of nucleation, domain growth and phase coexistence are required to provide the most convincing evidence of a first order transition, but these have not been collected for ferromagnets even though such a study has been performed for antiferromagnetic chromium[4] and changing domain structures obtained via spin reorientation have been noted for various ferromagnets[4,7,8].

$Fe_3Sn_2$ is a metallic ferromagnet with a lattice consisting of iron kagome bilayers separated by tin layers[9]. The Curie temperature is $T_c$=640 K and spin reorientation occurs upon cooling, where the spins rotate from close to the *c*-axis towards the *ab*-plane[10-13]. Even though it is agreed that the angle that the moments make with respect to the ab-plane is temperature-dependent, there is no report on the precise direction of the magnetic moment as a function of temperature. In this paper we assume that the angle that the moment makes with respect to the ab-plane is 90 degrees in the high temperature phase and 0 degrees in the low temperature phase.

Initial work on the spin reorientation transition in $Fe_3Sn_2$ was undertaken using Mössbauer spectroscopy on powder samples, where a transition was observed at 114 K[14]. Subsequent measurements showed that the spectra above 220 K consisted of one peak related to the population of spins in the high temperature phase whereas below 200 K two additional peaks characterizing the low temperature phase emerged, with the three peaks co-existing over a wide (80 K <T <220 K) range of temperatures[11]. The population of the low temperature (in-plane polarized) state was shown to grow as the temperature was lowered while that of the high temperature phase decreased. The spin rotation was suggested to be *abrupt* based on the constancy of the location of the peaks[11].

Subsequent measurements, again performed with powder samples, were associated with the proposal that the magnetic reorientation transition involves a *continuous* rotation of the spins from the *c*-axis towards the kagome plane[10,12]. Neutron powder diffraction data, where Bragg intensities were analysed, could be interpreted as following from either a continuous or abrupt spin rotation [10]. As a result, the nature of the spin reorientation, whether it is first or second order, has been open to debate.

Besides being magnetically interesting, $Fe_3Sn_2$ displays an exceptional anomalous Hall effect and an intrinsic mechanism was proposed that was due to the non-trivial magnetic spin texture experienced by the conduction electrons[15]. The material began attracting attention[16] as a candidate host for flat, topological two-dimensional bands analogous to the Landau levels in two-dimensional electron gases, and various anomalous electronic characteristics[17-19], including those associated with a Weyl semimetal state[20], have been reported. Research on single crystals where magnetic microscopy (Lorentz force) measurements were also performed and led to the claim that near room temperature there are magnetic textures and magnetotransport anomalies which could be related to skyrmionic bubbles[21].

The present work focuses on the spin reorientation in $Fe_3Sn_2$ single crystals using a combination of bulk magnetometry and local direct space imaging by variable temperature MFM[22,23]. We show direct images of spin reorientation where nucleation and growth of the new phase as well as phase coexistence are clearly observed. Due to the coupling between the magnetic and electronic degrees of freedom, the spin reorientation in $Fe_3Sn_2$ has direct implications for its electronic properties. The complex magnetic domain pattern reflects the electronic domain structure in the system.

**Results**

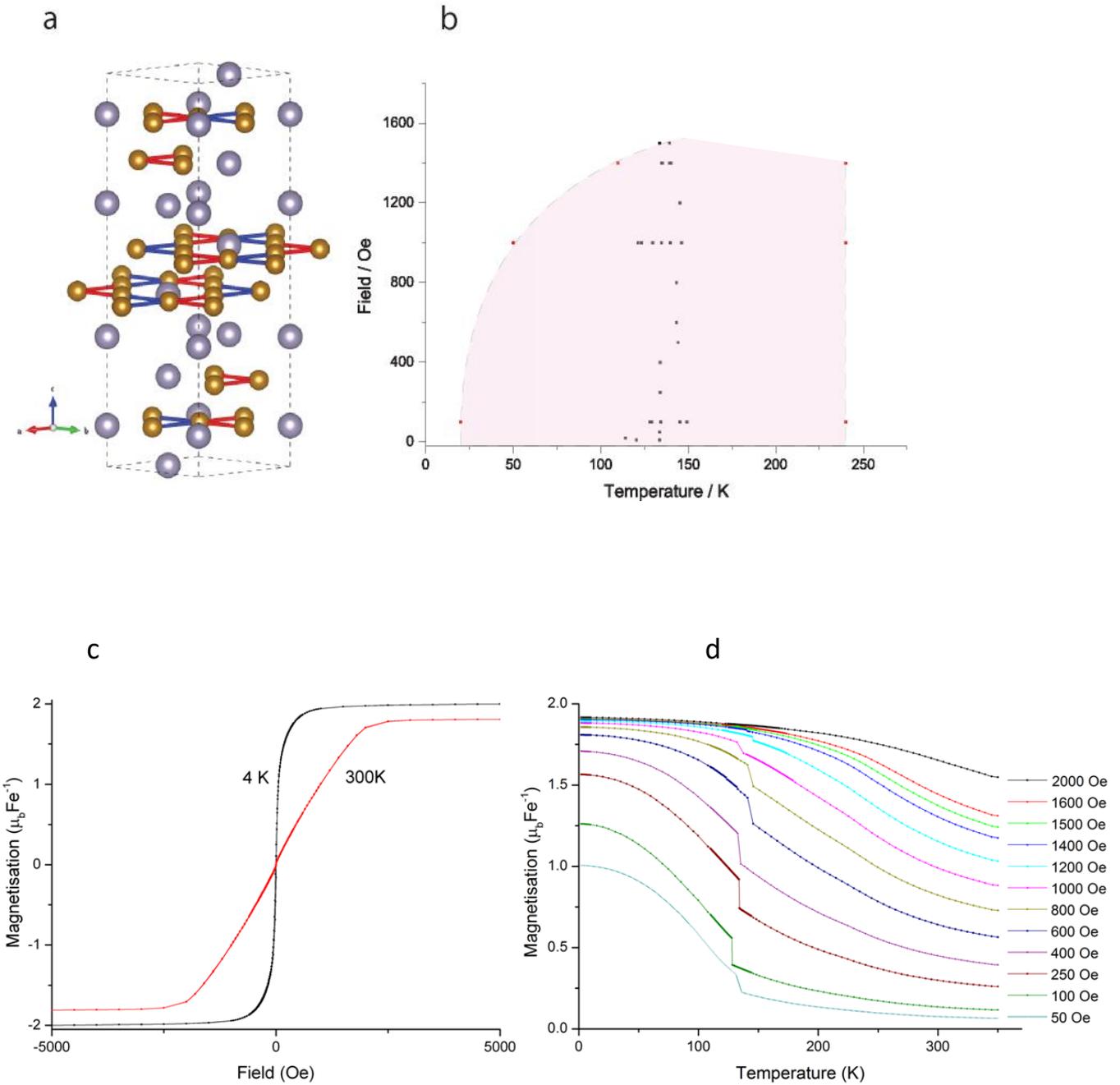

**Figure 1: (a) $Fe_3Sn_2$ crystal structure and (b) phase diagram showing the temperature and field of the coexistence of the high and low temperature phases (red points: boundary of the thermal hysteresis, black points: temperature of first order step). SQUID magnetometry data from crystal A (c) with the magnetic field applied along the *ab*-plane showing the change in magnetization at 300 K and 4 K and (d) temperature dependence of magnetization along the *ab*-plane on cooling at various magnetic fields showing the variation in the size of the first order step in magnetization.**

Figure 1a shows the structure; the equilateral triangles in the kagome planes have sides of length 2.732 and 2.582 Å[10]. The kagome planes exist in bilayers with a stacking period of 6.69 Å. Despite being metallic, there is a localized moment of 1.7 $\mu_b$ (300 K) to 1.9 $\mu_b$ (4 K) per iron atom. The magnetometry data in figure 1c,d show the change in magnetization within the *ab*-plane as a function of temperature and magnetic field for crystal-A (those from crystal B are shown in the Supplementary Materials). Crystal B was selected for MFM imaging due to its larger size. The data were collected both on cooling and warming with a measuring field between 50 and 2000 Oe in the *ab*-plane. The increase in magnetization on cooling indicates that the moments rotate from the *c*-axis towards the *ab*-plane, consistent with previous reports[Error! Bookmark not defined.,Error! Bookmark not defined.]. In addition, we observe large discrete jumps on cooling (but not on warming) at temperatures which appear to be randomly distributed near 130 K. Furthermore, we observe thermal hysteresis, *i.e.* the magnetization in the *ab*-plane is larger on warming than on cooling. In contrast, the field sweeps show no evidence of field hysteresis with both the high temperature (300 K) and low temperature phases (4 K) indicating that Fe$_3$Sn$_2$ is a soft magnet with negligible coercive field and no memory of field history. The variation in the field saturated magnetization is very small between 4 K and 300 K, indicating that the main effect of the temperature in the range of 4-350 K is to cause the reorientation of the moments.

We notice that at 300 K, where the easy axis is essentially perpendicular to the planes, approximately 2000 Oe are required to rotate the moment into the plane. Therefore, when the applied magnetic field in the *ab* plane is 2000 Oe, the moments are already rotated into the *ab* plane and thus there is negligible spin reorientation upon cooling, as shown in Fig 1d. The point (*T*=130 K, *H*=1500 G) in the temperature-field plane is analogous to the classic liquid-vapour critical point where the phase boundary between liquid and vapour disappears when the pressure is above the critical pressure. Concomitantly, the critical end point which we have discovered defines an external field beyond which single domain physics will determine the electron transport. Fig. 1b is the corresponding phase diagram where we draw the magnetic states as a function of temperature and field along the ab plane based on the magnetometry data. The pink region was defined by the points where the thermal hysteresis loops close at various magnetic field values and is where the high and low temperature phases coexist.

The magnetometry data from the aligned single crystal sample show discrete jumps that were not previously observed in powder samples as there they would be masked by powder averaging. For powders, where there is a uniform distribution of crystal orientations, the increase in the magnetization along the applied field observed by magnetometry upon cooling[10] could be accounted for by the higher probability that the *ab* plane rather than the *c*-axis of a crystallite is nearly parallel to the external field. Because there is no statistical sampling over crystallite orientations, the single crystal data highlight the thermal hysteresis and abrupt jumps in the magnetization on cooling as the moments rotate towards the *ab*-plane. We observe a variation in the size of the jumps between samples with most samples showing very small jumps.

To avoid the averaging of bulk magnetometry over domains, we investigated the underlying physics of the phase transition using MFM as a direct space local probe. The MFM data in figure 2a-f show the evolution of the domain structure on cooling from 200 to 4 K. Over this temperature range the evolution of the spin alignment from largely along the *c*-axis to towards the *ab*-plane changes the images drastically. As the contrast in the MFM images is due to stray fields perpendicular to the sample surface, the range of signals observed by MFM from the magnetic images decreases as the moments rotate towards the *ab*-plane. Consequently, the magnetic domain images at low temperatures show contributions from topography as well as magnetism, implying that steps (corresponding to the straight lines) in the surface structure are visible.

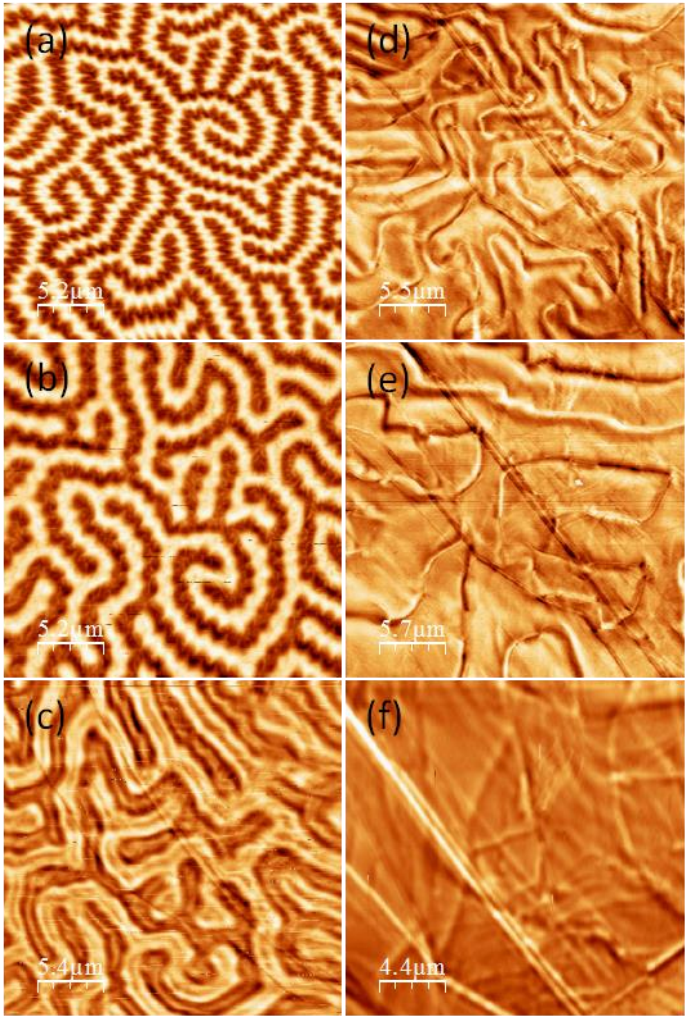

**Figure 2: MFM images showing the moments rotating from the out of plane direction to in-plane. The fine dendrites on the branched structure is lost on cooling: (a) 200 K (b) 160 K (c) 120 K (d) 100 K (e) 80 K (f) 4 K.**

The domain structure at 200 K is similar to that observed at room temperature and consists of a highly branched structure with fine dendritic fingers. The sharp contrast between the magnetic domains with magnetization pointing up or down indicates that the direction of the magnetic moment changes abruptly between the two domains. Such branched domain patterns are typical of highly anisotropic materials where the moments align perpendicular to the surface, and occur to minimise the stray field[24,25]. On cooling from 200 to 4.2 K, the spin structure changes from being primarily due to out-of-plane moments to consisting predominantly of in-plane domains. Correspondingly, the branching of the domain walls reduces, giving way to looping curves which become ever straighter. On further cooling the loops are not visible and domain walls associated with the low temperature phase appear.

Our data show that the initial stages of the domain structure evolution on cooling from high temperature lead to the development of a fragmented fine structure within the interior of the branched domains. This evolution is shown from 200 to 4 K in figure 2a-f and in greater detail from 160 to 140 K in figure 3a-c, with the images taken from the same sample area. The image sequence shows that between 160 and 140 K there is a growth of disjointed fine structures at the interior of the branches that have opposite contrast to the branch within which they reside. Initially, these fine filaments emanate from the tips of the nearly periodic dendritic protrusions on the walls between the bright and dark regions, i.e. they are nucleated at points of maximum field gradient. Further cooling leads to growth and extension of the fine structure throughout the branches and by 140 K the thin filaments are also largely parallel to rather than emanating from the major walls. The image at 120 K in figure 2c clearly shows that the inner filaments have become broader, continuous inner cores within both up and down branches.

Throughout the cooling process, there is a smoothing not only of the inner filaments but also of the principal walls between up and down domains, with the fine dendritic features that are easily seen at 160 K (Fig. 2b) being completely lost at 120 K (Fig. 2c). The smoothing and broadening follows from a temperature-dependent reduction of the anisotropy energy which confines the spins to the $z$ direction perpendicular to the kagome planes.

The inner core regions of the branches continue to expand on cooling; in the 100 K image (Fig. 2d), the outer edges of the branches separate due to the expansion of the inner core. The 4 K image (Fig. 2f) shows large domains with predominately in-plane magnetic moments and a number of low contrast magnetic domain walls.

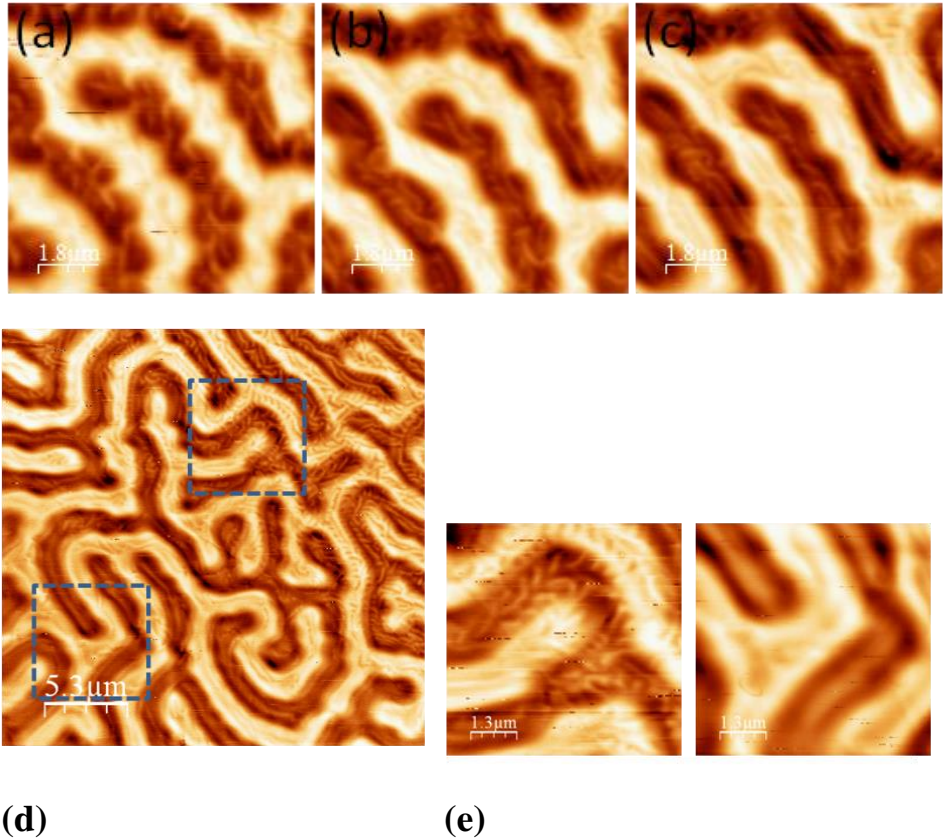

**(d)**             **(e)**

**Fig. 3 MFM images showing the development of the fine structure on cooling at (a) 160 K (b) 150 K and (c) 140 K. (d) MFM image at 130 K on cooling showing the coexistence of a high temperature (160 K) and low temperature (130 K) domain structure. (e) Separate scans of the regions inside the squares of (d)**

The subtle fine structure in figure 3d observed at 130 K suggests that there is a variation in the degree of transition, with the scan across the sample showing regions with different degrees of structure within the branches. These correspond to different stages of nucleation and growth through the spin reorientation transition. Domains in the top right of figure 3d have a fine structure, while in the bottom left far thicker core-like structures are present. The images in Fig. 3e correspond to separate scans of smaller regions. The radial average of their two-dimensional (2D) fast Fourier transform (FFT) shown in Fig. 4b can be used to characterise the contrasting length scales present in the images, and shows a larger high frequency component above 0.002 nm-1 in the top right quadrant compared to the bottom left quadrant. The radial average of the 2D-FFT at 160 and 120 K in Fig. 4b confirm that the high frequency component is associated with fine structure in the higher temperature (160 K) domain structure.

We observe the features characterizing the spin reorientation phase transition in more than one sample and on warming as well as cooling. In Fig. 4, we show MFM images for a different sample taken on warming from 77.73 K to 170 K. The images clearly show how the sample transitions from the low temperature phase to the high temperature phase. In the 77.7 K images there is a generally featureless in-plane domain structure with very little contrast. On warming to 120 K, the large featureless in-plane domain structure is lost and the whole volume consists of a meandering stripe (string) domain structure. What is interesting at 140 K is that the domain structure is built upon the string-dominated structure seen at 120 K but with hair-like side branches appearing. In other words, we are seeing the reverse of the process which we saw on cooling, where the filamentary "hair" disappears. Furthermore, while at 120 K the domain pattern can be described as an alternation of bright and dark regions with a periodicity slightly less than 1 micron (distance from bright to bright region), at 140 K we have another modulation of longer wavelength superimposed on the alternation observed at 120 K. The longer wavelength modulation produces an overall darker or brighter region, which we believe is the precursor of the high temperature phase, except that it has a smaller wavelength modulation inside. As we raise the temperature to 150 K and 160 K, the domain structure remains the same except for more addition of side branches. Between 160 K and 170 K, the morphology of the domain structure stays the same with the exception of a clearer distinction between overall bright vs dark regions.

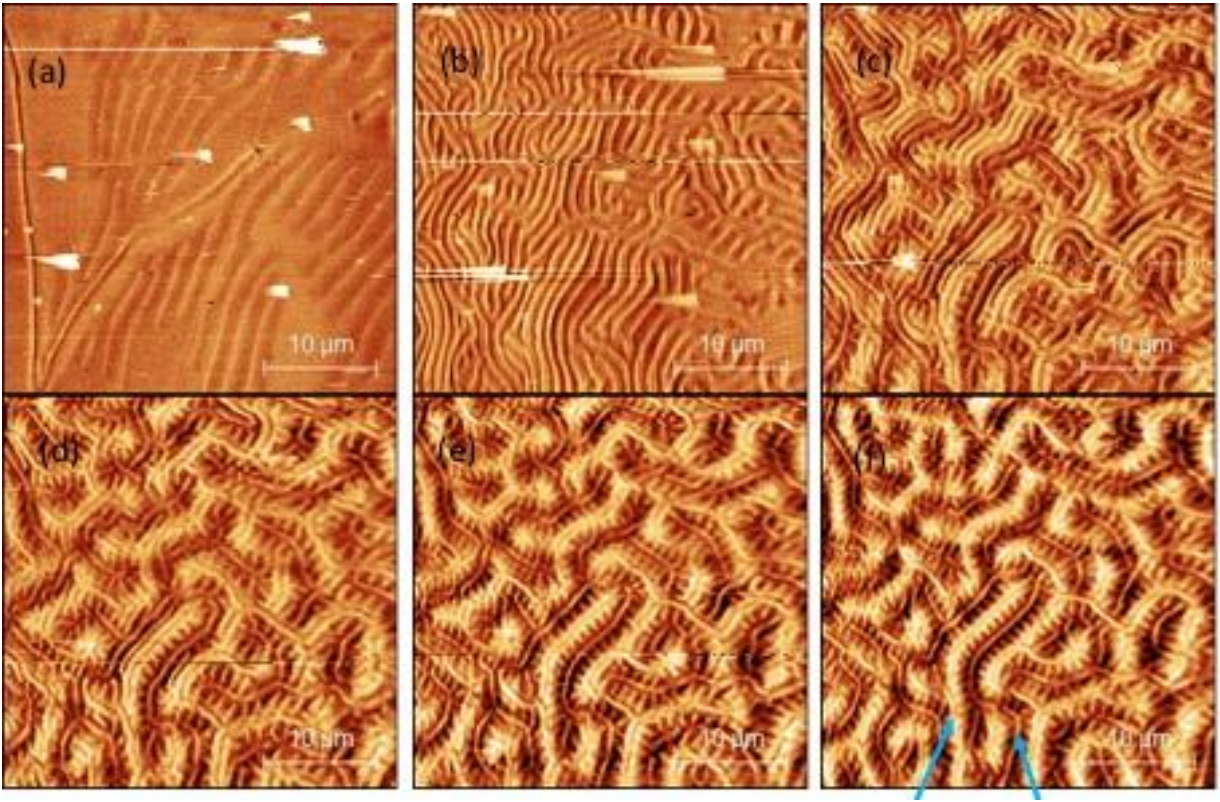

Fig 4. Images of the domain structure on warming. (a) 78 K, (b) 120 K, (c) 140 K, (d) 150 K, (e) 160 K, and (f) 170 K.

We believe that the strings marked by arrows for the 170 K image, either bright in the middle of a dark region, or dark in the middle of a white region are the same as the continuous inner cores that we observe on cooling. However, the temperature at which they disappear on warming is higher (T > 170 K) than the temperature at which they form on cooling (T < 140 K). The sign of the hysteresis of the inner core suggests that it belongs to the low temperature phase. Subsequent measurements on a different sample carried to higher temperatures show that the inner core is gone at 180 K.

MFM measures magnetic force along the *z* direction, and so when moments are predominantly in-plane, we expect a signal only near domain walls where the spin might rotate through *z*. This would imply passing from a strongly bimodal distribution for the MFM signal in the high-temperature phase, corresponding to up and down magnetic domains, to a single, narrower peak centred on zero that arises from in-plane magnetic domains. The histograms of the MFM frequency shifts (see Figure 5a) are in agreement with this expectation: there is a distribution which at high temperatures has two peaks which merge as the low temperature phase nucleates and increases in area. The merger occurs between 150 K and 100 K, the temperature range where large discrete jumps in the bulk magnetization are found to occur (Fig. 1d).

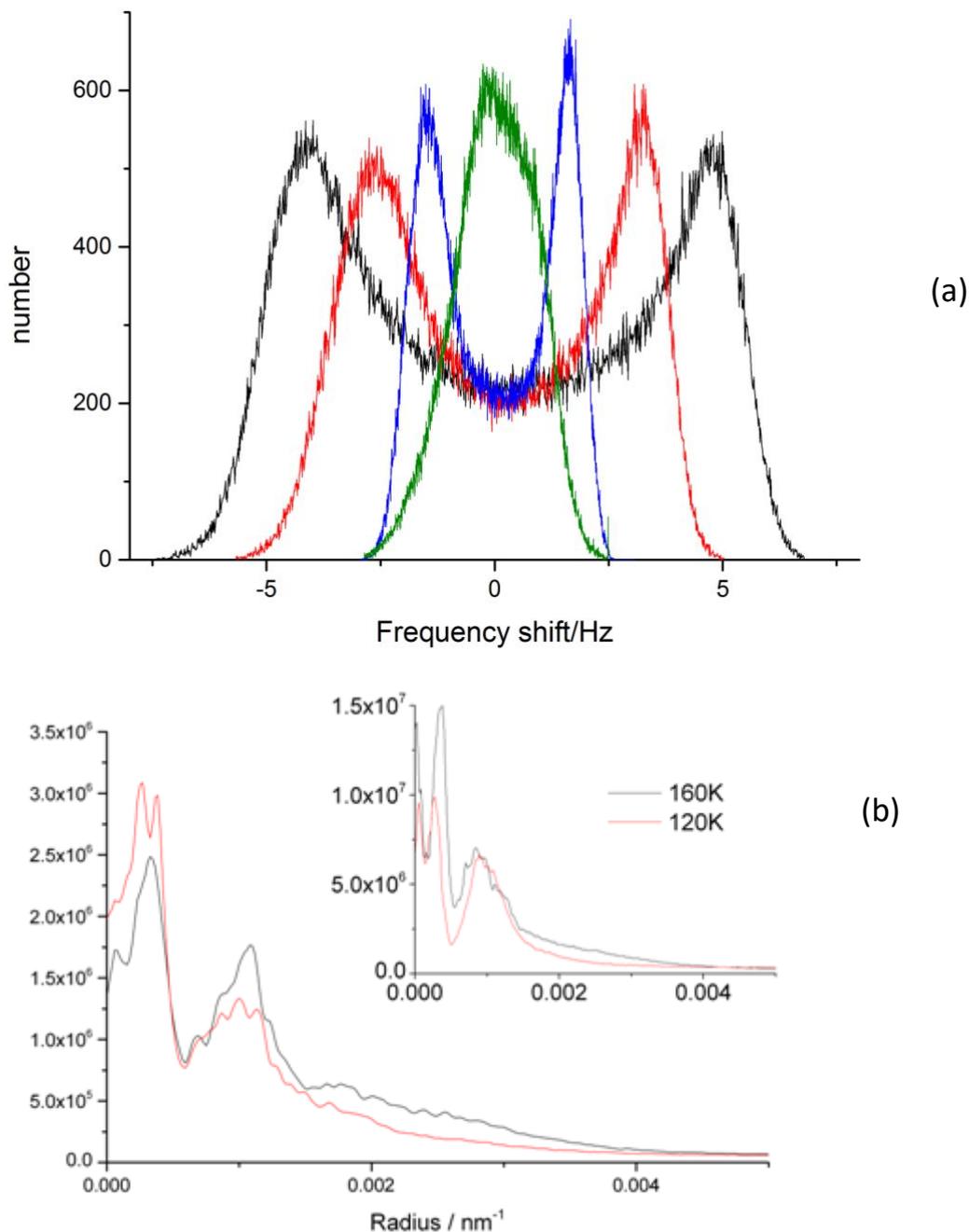

**Figure 5: (a) Histogram of the MFM images in Figure 2 showing the frequency shift associated with the domains. 200 K: black, 180 K: red, 150 K: blue, and 100 K: green. (b) Radial average of a FFT from the top right (black) and bottom left (red) quadrant of the MFM image in Figure 3 at 130 K, with the inset showing the radial average at 160 and 120 K.**

To understand the microscopic nature of the thermal hysteresis seen in bulk magnetometry, we compare MFM data, shown in figure 6, collected on cooling from 300 K and on warming from 80 K. The domain image at 150 K taken on cooling shows an extensive fine structure due to the formation of the low temperature phase and the subsequent growth of the in-plane phase. In contrast, the image collected at the same temperature after warming from 80 K shows a continuous inner core structure, indicating that the development of the high temperature phase on warming from 80 K follows a qualitatively different path than that leading to the low temperature structure upon cooling from 300 K, thus accounting for the long-observed thermal hysteresis of $Fe_3Sn_2$. The images at 150 K reveal that upon warming, vestiges of the low temperature phase remain as cores within the branches of the high temperature phase. Examination of Fig. 6, and in particular comparison of (b) and (c) also reveals that multiple warming and cooling cycles yield patterns with identical microstructures, but with numerous features on longer length scales rearranged.

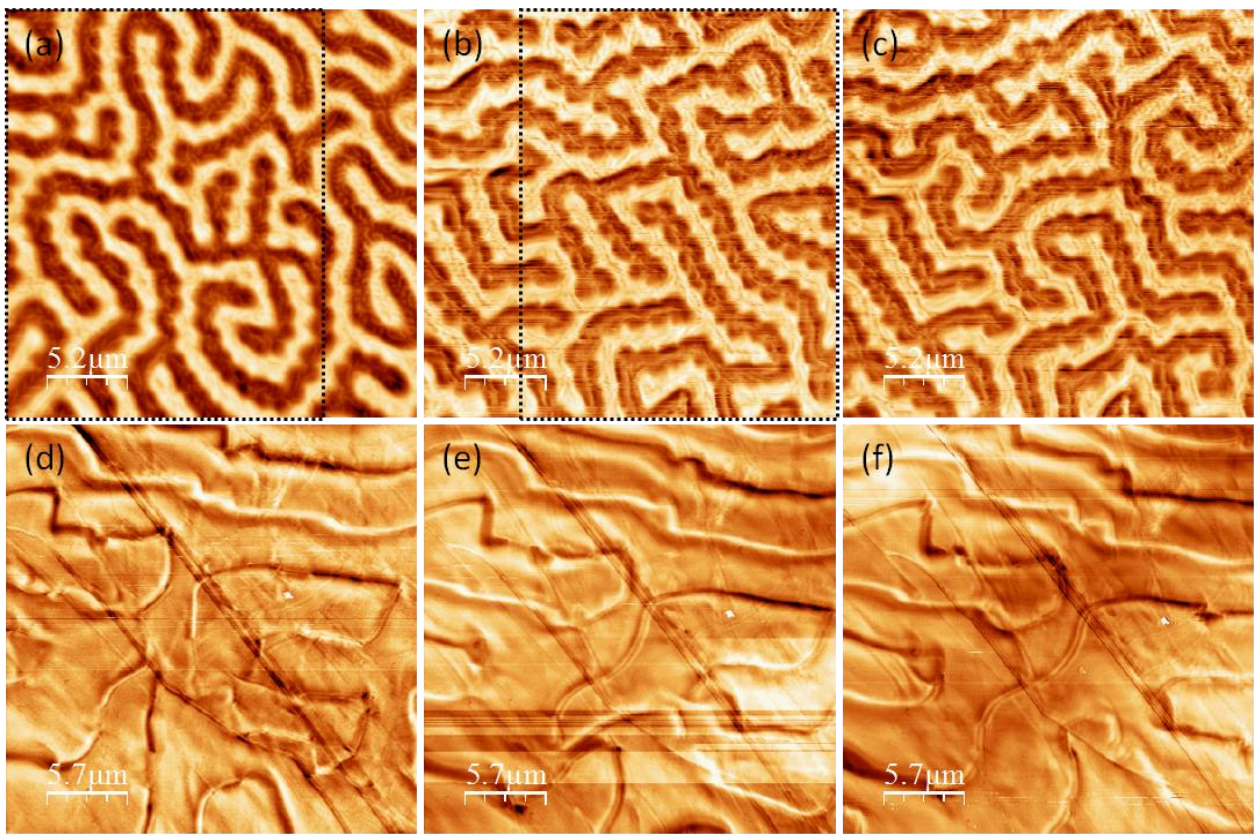

**Figure 6: MFM images of crystal B from cycling the temperature between 150 K (a-c) and 80 K (d-f) demonstrating thermal hysteresis in the domain structure.. The image sequence was (a), (d), (b), (e), (c) then (f) Image (a) at 150K was measured after cooling from 300K while images (b-c) were obtained on warming from 80K. There was no motion in the field of view of the microscope between the various data at 80K, but there was a slight shift between the data at 150K on the initial cool-down and first warm-up from 80K, which is indicated by the motion of the (dashed) bounding boxes between (a) and (b) .**

The presence of a continuous inner core structure persisting to higher temperature on warming is also observed on a different sample, which we studied in detail both on cooling and warming. The domain pattern of this sample is significantly more complex than of any other sample (Fig. 7). Even the high temperature phase (Fig. 7a), consists of more than alternating bright and dark domains. More recent measurements obtained by XMCD-PEEM on a different sample show a similar domain pattern at high temperature showing that this complex domain pattern is not an isolated case.

The images at left in Figure 7 were obtained during cooling, whereas the images at right were obtained upon warming to the same temperature as at left. At first instance, we clearly notice that the images on the left are different from those on the right, implying that the domain pattern on warming is different from that on cooling. On cooling we observe a substantial change of the domain pattern around 140 K becoming more obvious at 130 K. After careful examination of the images, what is thermally hysteretic besides the domains and domain walls having moved is in essence the presence of the continuous inner core structure characteristic of the low temperature phase up to higher temperatures during warming. As indicated by the blue lines, this inner core structure persists up to 160 K, but are not visible at 180 K.

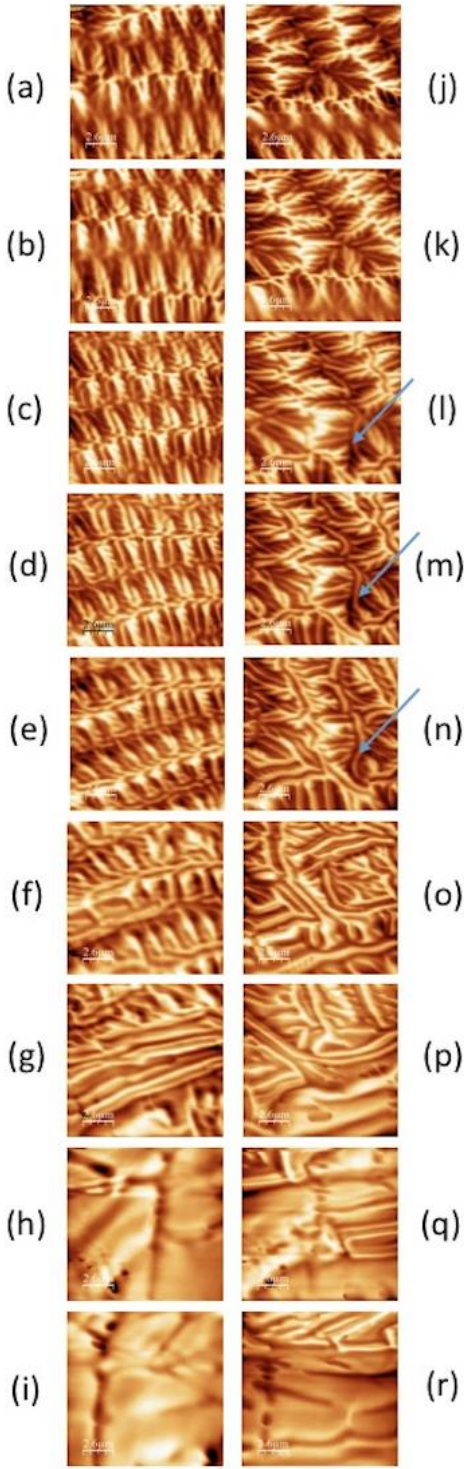

Figure 7. Images of domain structure taken during cooling (a)-(i) and warming (j)-(r) at various temperatures. (a, j) 200 K, (b, k) 180 K, (c, l) 160 K, (d, m)150 K, (e, n) 140 K, (f, o) 130 K, (g, p) 120 K, (h, q) 110 K, and (i, r) 100 K. During warming (l)-(n) from 140 K up to 160 K, we observe a string like structure indicated inside the blue ovals which we believe are of the same nature as those observed in Fig 4 reflecting the low temperature phase. As in Fig 4, they disappear at the same temperature and they are no longer observed at 180 K.

## Discussion

Our single crystal magnetic force microscopy and magnetometry data show that the much-studied kagome ferromagnet $Fe_3Sn_2$ undergoes a first order spin reorientation transition which in the magnetic field-temperature plane is analogous to a liquid gas transition in the pressure-temperature plane. The images reveal that ferromagnetic domain walls, as well as nucleation and growth of the minority phase on both cooling and warming are responsible for the hysteretic effects. The fact that hysteresis is only present when cycling the temperature and not the magnetic field indicates that the thermal hysteresis arises from the first order nature of the spin reorientation transition and not because of spin glass behaviour.

A key aspect of a first order phase transition is that the order parameter, in this case the magnetic easy axis, changes abruptly through the phase transition. Our experiments have resulted in the discovery of such jumps in certain small samples of $Fe_3Sn_2$, while also explaining the more gradual evolution of the magnetization in others on account of nucleation and growth of small domains.

The MFM images reveal the development of a fine magnetic structure on cooling that corresponds to the nucleation and growth of the low temperature phase from the domain walls of the high temperature phase. This is reminiscent of the growth of the longitudinal spin density wave state from the domain walls separating regions with different orientations of the spin density wavevector for the transverse phase in chromium[4]. Although the order parameter in chromium is more complex than for ferromagnetic $Fe_3Sn_2$, the reason for nucleation at high temperature phase domain walls is the same – at the domain walls, the magnetic moments can rotate into the direction favoured in the low temperature phase.

At zero magnetic field, the thermal hysteresis spans a large temperature window, from temperatures below 50 K to temperatures close to or above 200 K. In this temperature window, phase coexistence of magnetization along c and along the ab plane occur. We have shown that at temperatures around the midpoint of the thermal hysteresis temperature window, i.e. around 120 K, the two magnetic phases that coexist in a non-equilibrium system self-organize with the low temperature phase running through the core of the high temperature phase domains. This is a realization of self-organization of the spin degrees of freedom in a clean homogeneous system and can be contrasted to other examples of self-organization where external factors such as substrate bending[26] or polymer mixing[27] lead to self-organization.

The microscopic origin of the change in the preferred spin direction is also likely to be the same for $Fe_3Sn_2$ and Cr, namely a cooling-induced change in the band structure when spin-orbit effects are included[20]. More specifically, there are different band structures for the possible spin orientations[20], and as the chemical potential is changed with temperature, the free energy for one orientation crosses that for the other, resulting in a simple first order transition. For ferromagnetic $Fe_3Sn_2$ and not for antiferromagnetic Cr, the magnetic dipolar interaction also enters, resulting in a much broader transition, and especially when moments are perpendicular to the sample plane, much more complex and ramified domain walls.

Our results are significant not only in showing that we are dealing with different band structures that yield different free energies for different magnetization directions in $Fe_3Sn_2$, but also in revealing complex domain wall phenomena which are strongly temperature dependent. This implies that over a large temperature range, low field magnetotransport occurs in a highly electronically inhomogeneous ferromagnetic Weyl semimetal medium[20] where we are dealing with more than ordinary domain walls and skyrmions[28]. This means that we need to consider electrons moving across boundaries between thermal history-dependent regions which have different band structures near the Fermi level even though they are hosted by the same crystal.

**Materials and Methods**

$Fe_3Sn_2$ powder was prepared from stoichiometric quantities of Fe and Sn that were ground together and pressed into a pellet. This was placed into a silica ampoule and evacuated to $10^{-5}$ mbar before being back filled to 3.5 mbar with argon to reduce Sn evaporation. The ampoule was heated to 800°C at a ramp rate of 1°C/min and left for 7 days. The reaction was then quenched by submersion in cold water. Single crystals were prepared by chemical vapour transport from 500 mg of the prepared $Fe_3Sn_2$ powder loaded into a 16 cm long silica ampoule with 40 mg of iodine. The ampoule was evacuated to a pressure of $10^{-6}$ mbar and sealed. After heating at a rate of 1°C/min in a two-zone furnace to 650 and 720 °C, the reaction was left over 8 days. It led to crystals up to 6 mm in diameter.

A Quantum design MPMS dc-SQUID magnetometer measured the bulk properties of the single crystals, which were aligned with their *c*-axes both perpendicular and parallel to the applied magnetic field using a purpose-built sample holder where Apiezon-N grease was used for adhesion. The influence of the magnetic history of the crystals on the magnetometry was minimised by warming to 350 K in zero-field before measuring.

Direct space images of the magnetic domains were collected using an Attocube variable temperature MFM with the sample cleaved along the *ab*-plane prior to imaging. The samples and MFM system were held under an inert helium atmosphere at 1 mbar in a cryostat. Sample temperature was controlled by local counter-heating. A region of the sample with a few topographical features was selected for study, allowing the signal from the magnetic domains to be easily isolated from the topographical features while the latter allowed for tracking the same regions with changing temperature. We employed a Veeco MESP-LM MFM tip with a moment 0.3 $e_{-13}$ emu, with the moment of the tip oriented along the *c*-axis of the sample, using a lift height of 50 nm.   The magnetic images are a two-dimensional map of the shift in the resonance frequency of the cantilever with a magnetic tip, driven near resonance.

# Supplementary Materials

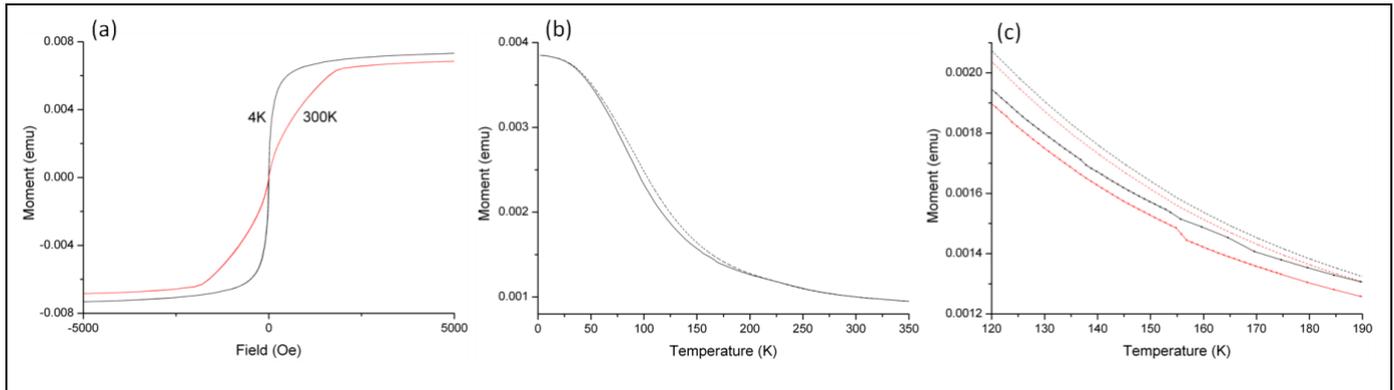

Supplementary figure 1: SQUID magnetometry data from crystal B with the magnetic field applied along the ab plane showing the change in magnetisation with (a) magnetic field and (b, c) temperature in a 100 Oe field (cooling - solid line, warming-dashed line). Plots (b, c) show small steps in magnetisation on cooling, the red and black lines in (c) corresponding to separate measurements.

# References

1  Isnard, O., Long, G. J., Hautot, D., Buschow, K. H. J. & Grandjean, F. A neutron diffraction and Mossbauer spectral study of the magnetic spin reorientation in Nd6Fe13Si. *J Phys-Condens Mat* **14**, 12391-12409, doi:Pii S0953-8984(02)39916-8

Doi 10.1088/0953-8984/14/47/313 (2002).

2  Yen, F. *et al.* Magnetic field effect and dielectric anomalies at the spin reorientation phase transition of GdFe3(BO3)(4). *Physical Review B* **73**, doi:ARTN 054435

10.1103/PhysRevB.73.054435 (2006).

3  Mizusaki, S. *et al.* Ferromagnetism and spin reorientation in Sm12Fe14Al5. *J Magn Magn Mater* **322**, L19-L24, doi:10.1016/j.jmmm.2009.12.021 (2010).

4  Evans, P. G., Isaacs, E. D., Aeppli, G., Cai, Z. & Lai, B. X-ray microdiffraction images of antiferromagnetic domain evolution in chromium. *Science* **295**, 1042-1045, doi:DOI 10.1126/science.1066870 (2002).

5  Binder, K. Theory of 1st-Order Phase-Transitions. *Rep Prog Phys* **50**, 783-859 (1987).

6  Kuz'min, M. D. On the gradual character of the first-order spin reorientation transition in DyFe11Ti. *Journal of Applied Physics* **88**, 7217-7222, doi:Pii [S0021-8979(91)06001-2]

Doi 10.1063/1.1327605 (2000).

7  Huang, J. *et al.* Magnetic state of La(1.36)Sr(1.64)Mn(2)O(7) probed by magnetic force microscopy. *Physical Review B* **77**, doi:ARTN 024405

10.1103/PhysRevB.77.024405 (2008).

8  Seifert, M. *et al.* Domain evolution during the spin-reorientation transition in epitaxial NdCo5 thin films. *New Journal of Physics* **15**, doi:Artn 013019

10.1088/1367-2630/15/1/013019 (2013).

9  Malaman, B., Roques, B., Courtois, A. & Protas, J. Crystal-Structure of Iron Stannide Fe3sn2. *Acta Crystallogr B* **32**, 1348-1351, doi:Doi 10.1107/S0567740876005323 (1976).

10  Fenner, L. A., Dee, A. A. & Wills, A. S. Non-collinearity and spin frustration in the itinerant kagome ferromagnet Fe3Sn2. *J Phys-Condens Mat* **21**, doi:Artn 452202

10.1088/0953-8984/21/45/452202 (2009).
11	Lecaer, G., Malaman, B. & Roques, B. Mossbauer-Effect Study of Fe3sn2. *J Phys F Met Phys* **8**, 323-336 (1978).
12	Le Caer, G., Malaman, B., Haggstrom, L. & Ericsson, T. Magnetic-Properties of Fe3sn2 .3. Sn-119 Mossbauer Study. *J Phys F Met Phys* **9**, 1905-1919 (1979).
13	Malaman, B., Lecaer, G. & Fruchart, D. Magnetic-Properties of Fe3sn2 .2. Neutron-Diffraction Study. *J Phys F Met Phys* **8**, 2389-2399, doi:Doi 10.1088/0305-4608/8/11/022 (1978).
14	Trumpy, G., Both, E., Djega-Mariadassou, C. & Lecocq, P. Mossbauer-effect studies of iron-tin alloys. *Phys Rev B-Solid St* **2**, 3477-3490, doi:DOI 10.1103/PhysRevB.2.3477 (1970).
15	Kida, T. *et al.* The giant anomalous Hall effect in the ferromagnet Fe3Sn2-a frustrated kagome metal. *J Phys-Condens Mat* **23**, doi:Artn 112205

10.1088/0953-8984/23/11/112205 (2011).
16	Tang, E., Mei, J. W. & Wen, X. G. High-temperature fractional quantum Hall states. *Phys Rev Lett* **106**, 236802, doi:10.1103/PhysRevLett.106.236802 (2011).
17	Ye, L. *et al.* Massive Dirac fermions in a ferromagnetic kagome metal. *Nature* **555**, 638-642, doi:10.1038/nature25987 (2018).
18	Yin, J. X. *et al.* Giant and anisotropic many-body spin-orbit tunability in a strongly correlated kagome magnet. *Nature*, doi:10.1038/s41586-018-0502-7 (2018).
19	Wang, Q., Sun, S., Zhang, X., Pang, F. & Lei, H. Anomalous Hall effect in a ferromagneticFe3Sn2single crystal with a geometrically frustrated Fe bilayer kagome lattice. *Physical Review B* **94**, doi:10.1103/PhysRevB.94.075135 (2016).
20	Yao, M. *et al.* Switchable Weyl nodes in topological Kagome ferromagnet Fe3Sn2. *ArXiv e-prints* (2018). <https://ui.adsabs.harvard.edu/ - abs/2018arXiv181001514Y>.
21	Hou, Z. *et al.* Observation of Various and Spontaneous Magnetic Skyrmionic Bubbles at Room Temperature in a Frustrated Kagome Magnet with Uniaxial Magnetic Anisotropy. *Adv Mater* **29**, doi:10.1002/adma.201701144 (2017).
22	Soh, Y. A., Aeppli, G., Mathur, N. D. & Blamire, M. G. Magnetic phase transitions studied by magnetic force microscopy. *J Magn Magn Mater* **226**, 857-859, doi:Doi 10.1016/S0304-8853(00)01060-X (2001).
23	Soh, Y. A., Aeppli, G., Mathur, N. D. & Blamire, M. G. Mesoscale magnetism at the grain boundaries in colossal magnetoresistive films. *Physical Review B* **63**, doi:ARTN 020402

DOI 10.1103/PhysRevB.63.020402 (2001).
24	Soh, Y. A. & Aeppli, G. Temperature-dependent magnetism in transition metal films observed by magnetic force microscopy and classical magnetometry. *Journal of Applied Physics* **85**, 4607-4609, doi:Doi 10.1063/1.370423 (1999).
25	Welp, U. *et al.* Magnetic anisotropy and domain structure of the layered manganite La1.36Sr1.64Mn2O7. *Physical Review B* **62**, 8615-8618, doi:DOI 10.1103/PhysRevB.62.8615 (2000).
26	Wu, J. Q. *et al.* Strain-induced self organization of metal-insulator domains in single-crystalline VO2 nanobeams. *Nano Lett.* **6**, 2313-2317, doi:10.1021/nl061831r (2006).
27	Bates, F. S. POLYMER-POLYMER PHASE-BEHAVIOR. *Science* **251**, 898-905, doi:10.1126/science.251.4996.898 (1991).
28	Kobayashi, K., Ominato, Y. & Nomura, K. Helicity-Protected Domain-Wall Magnetoresistance in Ferromagnetic Weyl Semimetal. *Journal of the Physical Society of Japan* **87**, 5, doi:10.7566/jpsj.87.073707 (2018).